\documentclass[
aps,
prx,
twocolumn,
superscriptaddress,
floatfix,
longbibliography,
nofootinbib
]{revtex4-1}
\usepackage[dvipdfmx]{graphicx}
\usepackage{dcolumn}
\usepackage{bm}
\usepackage{ulem} 
\usepackage{amsmath}
\usepackage{amssymb}
\usepackage{txfonts}
\usepackage{hyperref}
\usepackage{color} 
\usepackage{xcolor}
\usepackage{listings}
\usepackage{cprotect}
\usepackage{bbm}
\usepackage{mathrsfs}
\usepackage{braket}
\usepackage{physics}

\graphicspath{{./figures/}}

\def\tcyan{\textcolor{cyan}}

\hypersetup{
  colorlinks=true,
  linkcolor=[rgb]{0.60,0.00,0.00},
  citecolor=[rgb]{0.00,0.00,0.60},
  urlcolor=[rgb]{0.00,0.00,0.60},
  setpagesize=false
}
\def\avg#1{\hat{#1}_{\rm avg}}

\makeatletter
\newsavebox{\@brx}
\newcommand{\vvert}[1][]{\savebox{\@brx}{\(\m@th{#1\vert}\)}%
 \mathclose{\copy\@brx\kern-0.5\wd\@brx\usebox{\@brx}}}
\makeatother

\begin{document}
\title{
Quantum synchronization and chimera states in a programmable quantum many-body system
}
\author{Kazuya Shinjo}
\affiliation{Computational Quantum Matter Research Team, RIKEN Center for Emergent Matter Science (CEMS), Wako, Saitama 351-0198, Japan}
\author{Kazuhiro Seki}
\affiliation{Quantum Computational Science Research Team, RIKEN Center for Quantum Computing (RQC), Wako, Saitama 351-0198, Japan}
\author{Seiji Yunoki}
\affiliation{Computational Quantum Matter Research Team, RIKEN Center for Emergent Matter Science (CEMS), Wako, Saitama 351-0198, Japan}
\affiliation{Quantum Computational Science Research Team,
RIKEN Center for Quantum Computing (RQC), Wako, Saitama 351-0198, Japan}
\affiliation{Computational Materials Science Research Team,
RIKEN Center for Computational Science (R-CCS), Kobe, Hyogo 650-0047, Japan}
\affiliation{Computational Condensed Matter Physics Laboratory,
RIKEN Cluster for Pioneering Research (CPR), Saitama 351-0198, Japan}

\date{\today}
 
\begin{abstract}
Synchronization is a hallmark of collective behavior in classical nonlinear systems, yet its realization as a robust many-body phenomenon in coherent quantum systems remains largely unexplored. Here we demonstrate symmetry-protected quantum synchronization and a quantum chimera state in coherent Floquet dynamics on programmable superconducting quantum processors.
By implementing stroboscopic evolution of a two-dimensional Heisenberg model on IBM heavy-hex devices, we observe that initially phase-randomized spins spontaneously self-organize into coherent lattice-wide oscillations. On 28 qubits, synchronization persists even for strongly randomized initial states and is stabilized by SU(2) symmetry, as confirmed by explicit symmetry breaking.
Scaling up to 156 qubits reveals a qualitatively distinct regime. For weak initial randomness, global synchronization extends across the device. For strong randomness, the system fails to synchronize globally, yet subsets of qubits exhibit robust local phase coherence under homogeneous unitary dynamics. This coexistence of globally desynchronized and locally synchronized regions constitutes a quantum analogue of a classical chimera state.
Statevector and matrix-product-state simulations reproduce both the symmetry-protected synchronization and the chimera coexistence, demonstrating that these phenomena arise from the intrinsic Floquet many-body dynamics.
Our results establish symmetry-protected synchronization and quantum chimera states as experimentally accessible nonequilibrium dynamical phases in programable many-body quantum systems.
\end{abstract}
\maketitle

%

\section{Introduction}

Collective order can emerge far from equilibrium, producing dynamical phases with no equilibrium analogue. Understanding such behavior in interacting quantum systems is a central challenge of modern many-body physics, particularly in two dimensions, where real-time evolution typically generates rapid entanglement growth. 
Although tensor-network methods have achieved remarkable success in describing ground states and short-time dynamics, their applicability to real-time evolution in two dimensions is fundamentally limited by entanglement growth and rapidly increasing computational cost~\cite{Orus2019,Weimer2021,Cirac2021}. 
This limitation motivates complementary experimental and computational approaches capable of accessing large-scale nonequilibrium dynamics in regimes that remain difficult to explore using classical simulation alone.

Synchronization is a paradigmatic form of dynamical order~\cite{Winfree1980,Elowitz2000,Pantaleone2002,Strogatz2005}. 
In classical systems, it describes the spontaneous phase alignment of coupled oscillators and is characterized by order parameters quantifying global coherence~\cite{Kuramoto1984,Strogatz2000,Acebron2005,Pikovsky2001}.
In interacting quantum systems, synchronization can emerge in the absence of classical limit cycles and is shaped by coherence, entanglement, and quantum fluctuations. 
Although signatures of quantum synchronization have been reported across diverse experimental platforms~\cite{Tindall2020,Vorndamme2021,Buvca2022,Reimann2023,Wadenpfuhl2023,Wu2024,Ding2024,Tao2025}, direct evidence of synchronized self-organization and its stabilizing mechanism in large, interacting two-dimensional systems under coherent many-body dynamics remains scarce.

One particularly striking synchronization pattern in classical nonlinear systems is the chimera state, in which coherent and incoherent oscillatory regions coexist despite homogeneous coupling and identical constituents~\cite{Kuramoto2002,Abrams2004}. 
Originally identified in networks of coupled phase oscillators, chimera states have since been observed in a wide range of classical settings and are now recognized as a fundamental example of emergent spatial heterogeneity in dynamical systems~\cite{Tinsley2012, Hagerstrom2012, Martens2013, Panaggio2015}. 
Whether analogous coexistence can arise intrinsically in closed quantum many-body systems remains unresolved. Unlike classical oscillator networks, quantum dynamics is governed by unitary evolution, where coherence, entanglement, and many-body interference reshape collective behavior. Demonstrating a quantum chimera state therefore requires establishing the spatial coexistence of synchronized and unsynchronized dynamics that originates from coherent many-body evolution rather than from engineered dissipation or external noise.

\begin{figure*}[htbp]
\includegraphics[width=1\textwidth]{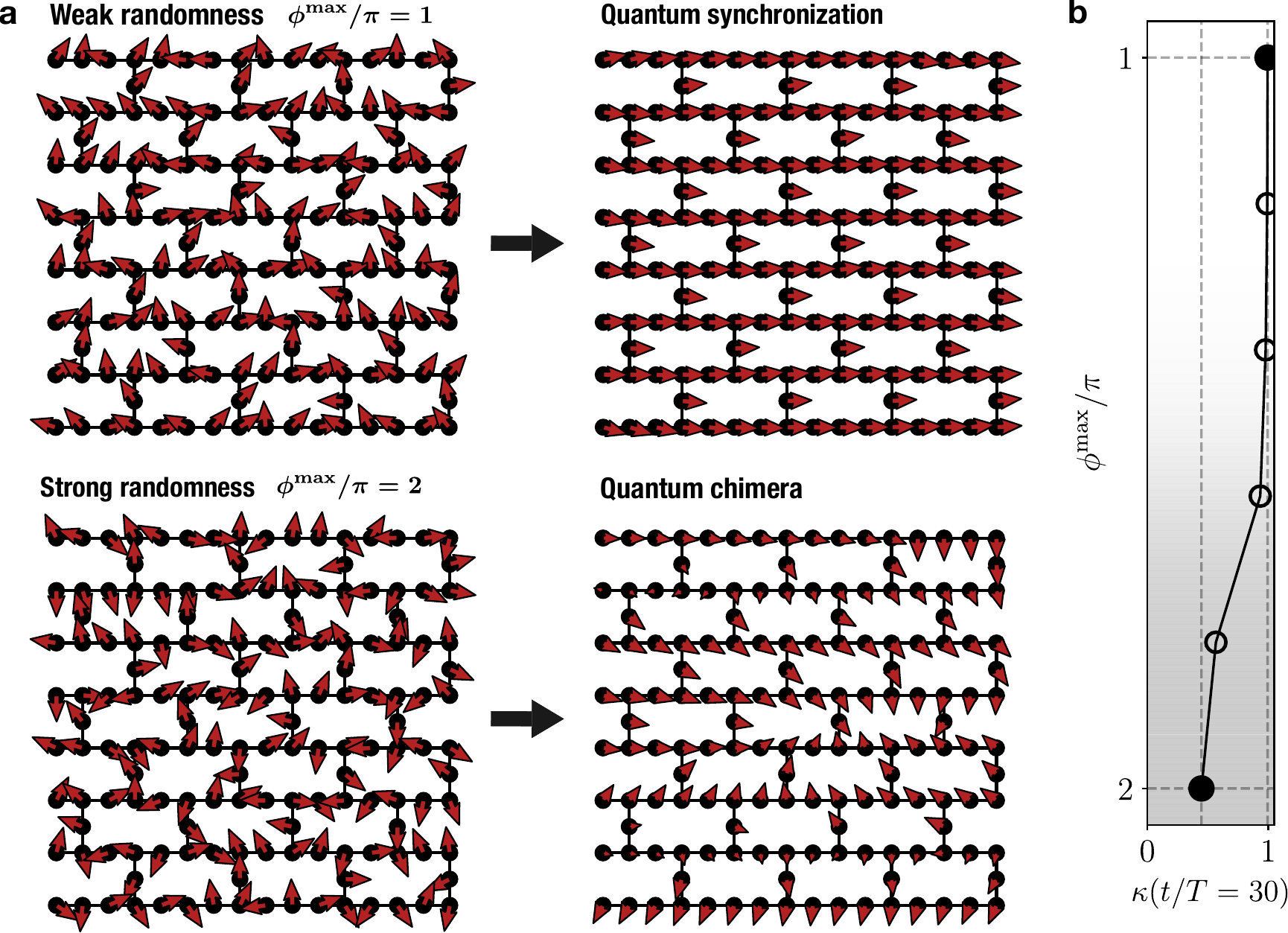}
    \caption{
    \textbf{Emergence of global synchronization and quantum chimera in a 156-qubit Floquet system.}
    \textbf{a,} Local in-plane magnetization vector field $(\langle \hat X_j\rangle,\langle \hat Y_j\rangle)$ from MPS simulations with bond dimension $\chi=600$ on a heavy-hex lattice with $L=156$.
    For weak phase randomness $(\phi^{\rm max}/\pi=1)$, an initially disordered configuration (left) evolves into a globally synchronized state at $t/T=30$ (right).
    For strong randomness $(\phi^{\rm max}/\pi=2)$, the system instead develops a quantum chimera at $t/T=30$ (right), where spatially separated regions retain internal phase coherence while remaining mutually incoherent. 
    \textbf{b,} Synchronization order parameter $\kappa$ at $t/T=30$ as a function of initial-state randomness $\phi^{\rm max}$. $\kappa=1$ corresponds to perfect global phase alignment, whereas $\kappa=0$ indicates complete absence of coherence; intermediate values reflect partial synchronization. 
    Black circles indicate the two values of $\phi^{\rm max}$ shown in \textbf{a}.
    Parameters are $(\theta_{XX},\theta_{ZZ},\theta_{Z})=(-0.25\pi,-0.25\pi,0.25\pi)$ (see Methods).
    Arrow lengths are rescaled independently in each panel for clarity.
    }
\label{fig:snapshots}
\end{figure*}

Here we use IBM heavy-hex quantum processors to investigate Floquet-engineered quench dynamics of the isotropic Heisenberg model in a longitudinal field. We observe the emergence of synchronized oscillations in the local magnetization and identify SU(2) symmetry as the stabilizing mechanism, confirmed by explicitly breaking this symmetry via anisotropy. 
On $L=28$ qubits, in-plane spins initialized with random phases dynamically align and sustain coherent oscillations, with error-mitigated measurements in quantitative agreement with statevector simulations. Extending to $L=156$ qubits reveals a qualitative change: increasing the degree of initial state randomness leads to the spatial coexistence of synchronized and desynchronized regions, constituting a quantum chimera state in analogy with its classical counterpart~\cite{Kuramoto2002,Abrams2004}. 
Matrix-product-state (MPS) simulations reproduce both the symmetry-stabilized synchronization and the chimera coexistence, demonstrating that these phenomena arise from the intrinsic Floquet many-body dynamics. These results establish symmetry-protected synchronization and quantum chimera states as robust nonequilibrium dynamical phases realized in coherent two-dimensional quantum systems.

\section{Results}\label{sec:4}

We investigate stroboscopic Floquet dynamics of the isotropic Heisenberg model in a longitudinal field (see Methods) implemented on IBM heavy-hex quantum processors. 
Our central observables are the local transverse magnetizations $\langle \hat X_j(t)\rangle$ and their phase coherence. 
We first establish symmetry-protected synchronization on $L=28$ qubits and then examine its behavior at larger system size. 
We further show that explicit breaking of SU(2) symmetry via anisotropy destroys synchronization. 
Figure~\ref{fig:snapshots} summarizes the dynamics for $L=156$ obtained from MPS simulations. For weak initial in-plane phase randomness, the system evolves into a globally synchronized state characterized by uniform phase alignment across the lattice. 
For stronger randomness, global synchronization fails to develop; instead, spatially separated regions maintain internal phase coherence while remaining mutually incoherent, constituting a quantum chimera state. 
Corresponding $L=156$ quantum hardware results are shown in Supplementary Fig.~S11.

\subsection{Symmetry-protected synchronization on a 28-qubit lattice}\label{sec:4sync}

\begin{figure}[htbp]
\includegraphics[width=0.48\textwidth]{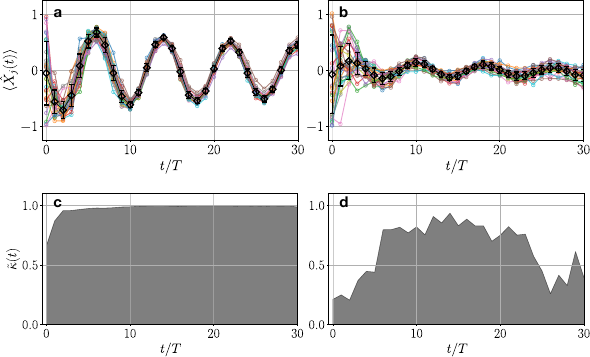}
\caption{
	\textbf{Symmetry-protected synchronization on a 28-qubit heavy-hex lattice.}
    Data are obtained from experiments on the \texttt{ibm\_kobe} quantum processor. 
    \textbf{a,} Error-mitigated local transverse magnetizations $\langle \hat X_j(t)\rangle$ for all qubits (colored circles) on the $L=28$ heavy-hex subgraph.
    Black diamonds show the spatial average $\langle \avg{X}(t)\rangle$, with error bars indicating the spatial standard deviation $X_{\rm std}(t)$. 
    The initial state has weak phase randomness, $\phi^{\rm max}/\pi=1$.
    \textbf{b,} Same quantities as in \textbf{a} for strong phase randomness, $\phi^{\rm max}/\pi=2$. 
    \textbf{c,} Synchronization order parameter $\tilde{\kappa}(t)$ computed from the data in \textbf{a} (see Supplementary Information Sec.~S1).
    \textbf{d,} $\tilde{\kappa}(t)$ computed from the data in \textbf{b}.
    Parameters are $(\theta_{XX},\theta_{ZZ},\theta_Z)=(-0.25\pi,-0.25\pi,0.25\pi)$.
	}
\label{fig:globalSyncL28}
\end{figure}

We begin with a $L=28$ heavy-hex subgraph implemented on \texttt{ibm\_kobe} (region enclosed in Fig.~\ref{fig:circuit}a).
Figures~\ref{fig:globalSyncL28}a,b display the error-mitigated dynamics of local transverse magnetizations $\langle \hat X_j(t)\rangle$ for $(\theta_{XX},\theta_{ZZ},\theta_{Z})=(-0.25\pi,-0.25\pi,0.25\pi)$.
Unless otherwise stated, all quantum hardware expectation values shown in this work are error mitigated using the protocol described in Eq.~(\ref{eq:method-mitigation}) (see Methods and Supplementary Information Sec.~S3). 
The initial state is a product state with local Bloch vectors randomized in the $xy$ plane, controlled by $\phi^{\max}$ (see Methods). 
Starting from disordered phases at $t=0$, the spins dynamically self-organize into a regime where oscillations of $\langle \hat X_j(t)\rangle$ become phase-aligned across the lattice, signaling the emergence of collective synchronization. 
The spatially averaged magnetization $\langle \avg{X}(t)\rangle$ exhibits coherent oscillations, while the spatial dispersion $X_{\rm std}(t)$ is strongly suppressed in the synchronized regime.
Here, $X_{\rm std}(t)$ denotes the spatial standard deviation of $\langle \hat X_j(t)\rangle$ over $j\in M$, with $M=\{0,1,2,\dots, 27\}$ (Supplementary Information Sec.~S1.1).

To quantify phase alignment, we introduce a synchronization order parameter. 
Because $\langle \hat Y_j(t)\rangle$ cannot be measured simultaneously with $\langle \hat X_j(t)\rangle$ without additional overhead, we employ a hardware-efficient proxy $\tilde{\kappa}(t)$, obtained from phases extracted from $\langle \hat X_j(t)\rangle$ via the Hilbert transform~\cite{Gabor1946, Pikovsky2001} (see Supplementary Information Sec.~S1 for definition and benchmarking). 
For weak randomness $\phi^{\max}=\pi$, $\tilde{\kappa}(t)$ approaches unity, indicating near-global phase coherence (Fig.~\ref{fig:globalSyncL28}c). 
For strong randomness $\phi^{\max}=2\pi$, $\tilde{\kappa}(t)$ remains lower, reflecting reduced coherence for more strongly randomized initial states (Fig.~\ref{fig:globalSyncL28}d).

To assess the role of device noise, we perform exact statevector simulations for $L=28$ (Supplementary Figs.~S4a--f).
In the ideal limit, phase synchronization becomes nearly perfect for both $\phi^{\rm max}/\pi=1$ and 2, with $\kappa(t)\to 1$, and oscillation amplitudes align across qubits, such that $R_{\rm std}(t)\to 0$ while $R_{\rm avg}(t)$ remains finite.
Here $R_{\rm avg}(t)$ and $R_{\rm std}(t)$ denote the spatial mean and standard deviation of $R_j(t)=|\langle \hat X_j(t)\rangle+i\langle \hat Y_j(t)\rangle|$ over $j\in M$, with $M=\{0,1,2,\dots,27\}$ (Supplementary Information Sec.~S1.1).
Additional geometry-dependent analyses, including square and triangular lattices, based on statevector simulations are presented in Supplementary Information Sec.~S4. These results support quantum synchronization as an intrinsic nonequilibrium phenomenon in two-dimensional coherent many-body dynamics.

\subsection{Quantum chimera state on a 156-qubit lattice} \label{sec:4chimera}

\begin{figure*}[htbp]
\includegraphics[width=1\textwidth]{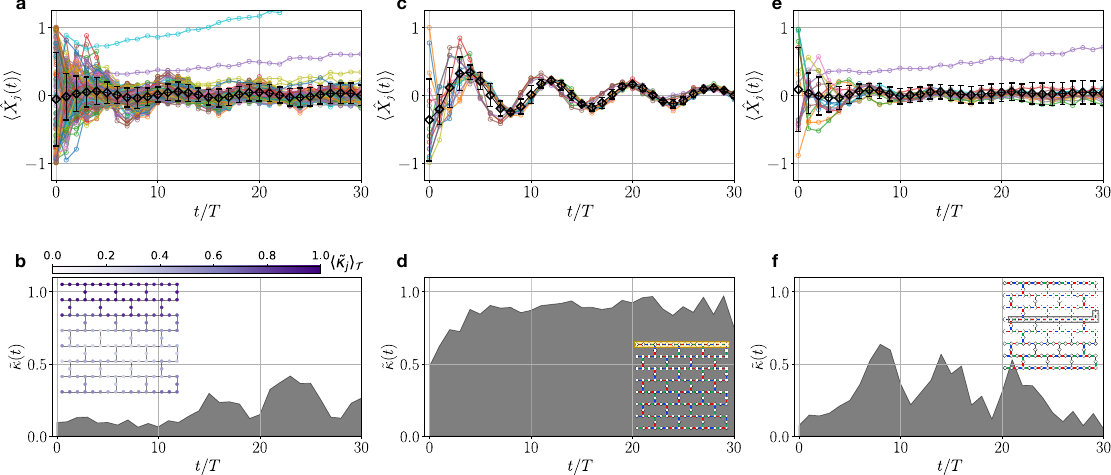}
\caption{
	\textbf{Quantum chimera dynamics on a $L=156$ heavy-hex lattice observed on the \texttt{ibm\_kobe} quantum processor.}
    \textbf{a,} Time evolution of the local magnetizations $\langle \hat X_{j}(t)\rangle$ for all qubits (colored circles).
    Black diamonds denote the spatial average $\langle \avg{X}(t)\rangle$ over all qubits $M=\{0,1,\ldots,L-1\}$, with error bars indicating the spatial standard deviation $X_{\rm std}(t)$. Two qubits ($j=64$ and 129) exhibit enhanced noise, but all qubits are retained in the analyses. 
    \textbf{b,} Global synchronization order parameter $\tilde \kappa(t)$ computed from the full data in \textbf{a}.
    Inset: spatial map of the late-time-averaged local synchronization order parameter $\langle \tilde \kappa_j(t)\rangle_{\mathcal{T}}$ on the heavy-hex lattice, evaluated within a fixed-$K$ neighborhood ($K=4$ qubits) around each qubit $j$ and averaged over time interval $\mathcal{T}=[20,25]$ (see Methods). 
    \textbf{c,d,} Time evolution of the local magnetizations $\langle \hat X_{j}(t)\rangle$ (colored circles) and the corresponding $\tilde{\kappa}(t)$ evaluated on qubits $M=\{0,1,\ldots,15\}$ highlighted in panel \textbf{d}, representing a synchronized subsystem.
    \textbf{e,f,} Same quantities as in \textbf{c,d} but for qubits $M=\{59,61,62,\ldots,75\}$ highlighted in panel \textbf{f}, representing a desynchronized region.
    All measured magnetizations are error-mitigated using the protocol described in Eq.~(\ref{eq:method-mitigation}). 
    Parameters are $(\theta_{XX},\theta_{ZZ},\theta_{Z})=(-0.25\pi,-0.25\pi,0.25\pi)$, and the initial state is prepared with strong randomness $\phi^{\rm max}=2\pi$.
}
\label{fig:L156_kobe_rand0pi2pi}
\end{figure*}

We next scale the system to $L=156$ qubits on the full heavy-hex lattice (Fig.~\ref{fig:circuit}a). 
For strong randomness $\phi^{\max}=2\pi$, global synchronization no longer develops: the spatial dispersion $X_{\rm std}(t)$ does not relax to zero (Fig.~\ref{fig:L156_kobe_rand0pi2pi}a), and the global order parameter $\tilde{\kappa}(t)$ remains small  (Fig.~\ref{fig:L156_kobe_rand0pi2pi}b), indicating the absence of global synchronization. 
At the same time, the lack of global order coexists with strong local coherence. 
The inset of Fig.~\ref{fig:L156_kobe_rand0pi2pi}b maps the late-time-averaged local coherence $\langle \tilde{\kappa}_{j}\rangle_\mathcal{T}$, obtained from phases extracted from $\langle \hat X_j(t)\rangle$ within a $K=4$ neighborhood (see Methods). 
Restricting to a 16-qubit subsystem, synchronized oscillations of $\langle \hat X_j(t)\rangle$ emerge with $\tilde{\kappa}(t)\to 1$ (Fig.~\ref{fig:L156_kobe_rand0pi2pi}c,d).
In contrast, another 16-qubit subsystem exhibits weaker coherence and $\tilde{\kappa}(t)$ remains small (Fig.~\ref{fig:L156_kobe_rand0pi2pi}e,f).
Thus, under a homogeneous Hamiltonian and uniform driving, synchronized and desynchronized regions coexist in space. 
Such coexistence is the defining feature of a chimera state~\cite{Kuramoto2002,Abrams2004} in classical systems, and we identify this regime as a quantum chimera state.

\begin{figure*}[htbp]
\includegraphics[width=1\textwidth]{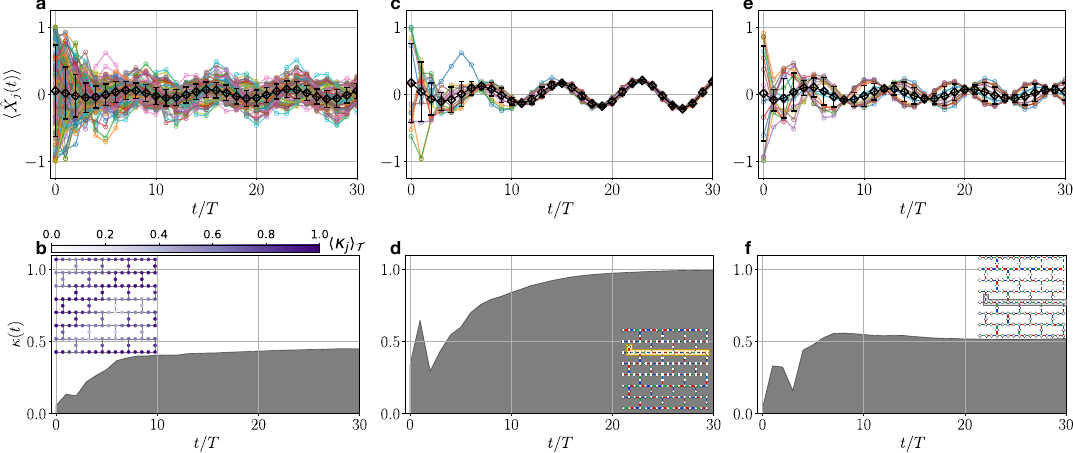}
\caption{
    \textbf{Quantum chimera dynamics reproduced by MPS simulations on the same $L=156$ heavy-hex system.}
    \textbf{a,} Time evolution of the local magnetizations $\langle \hat X_{j}(t)\rangle$ for all qubits (colored circles).
    Black diamonds denote the spatial average $\langle \avg{X}(t)\rangle$ over all qubits $M=\{0,1,\ldots,L-1\}$, with error bars indicating the spatial standard deviation $X_{\rm std}(t)$.
    \textbf{b,} Global synchronization order parameter $\kappa(t)$ computed from the full data in \textbf{a}.
    Inset: spatial map of the late-time-averaged local synchronization order parameter $\langle \kappa_j(t)\rangle_{\mathcal{T}}$, evaluated within a fixed-$K$ neighborhood ($K=4$ qubits) around each qubit $j$ and averaged over time interval $\mathcal{T}=[20,25]$ (see Methods). 
    \textbf{c,d,} Time evolution of the local magnetizations $\langle \hat X_{j}(t)\rangle$ (colored circles) and the corresponding $\kappa(t)$ evaluated on qubits $M=\{36,41,42,\ldots,55\}$ highlighted in panel \textbf{d}, representing a synchronized subsystem. 
    \textbf{e,f,} Same quantities as in \textbf{c,d} but for qubits $M=\{76,81,82,\ldots,95\}$ highlighted in panel \textbf{f}, representing a desynchronized subsystem.
    MPS simulations are performed with bond dimension $\chi=600$.
    Parameters are $(\theta_{XX},\theta_{ZZ},\theta_{Z})=(-0.25\pi,-0.25\pi,0.25\pi)$, and the initial state is prepared with strong randomness $\phi^{\rm max}=2\pi$.
}
\label{fig:L156_MPS_rand0pi2pi}
\end{figure*}

To assess whether the observed behavior originates from device noise, we perform MPS simulations of the same $L=156$ heavy-hex dynamics. 
The simulations reproduce the qualitative features observed in the experiment: global synchronization is absent (Fig.~\ref{fig:L156_MPS_rand0pi2pi}a,b), yet certain subsystems display near-perfect synchronization with $\kappa(t)$ evaluated within these subsystems approaching unity (Fig.~\ref{fig:L156_MPS_rand0pi2pi}c,d), while others remain only weakly coherent (Fig.~\ref{fig:L156_MPS_rand0pi2pi}e,f). 
Furthermore, weakly coherent regions can contain smaller synchronized pockets when examined at finer spatial scales, as illustrated in Supplementary Information Sec.~S6.3 and Figs.~S12 and S13. 
This hierarchical structure is a characteristic feature of \tcyan{quantum} chimera states, observed here in coherent dynamics of a closed quantum system. 
Convergence with bond dimension is verified in Supplementary Information Sec.~S5.2, supporting the interpretation that the coexistence is intrinsic to the model rather than a device-specific artifact.

For weak randomness $\phi^{\rm max}=\pi$, global synchronization is observed in both quantum hardware experiments and MPS simulations, as discussed in Supplementary Information Sec.~S6.1. 
Thus, the emergence of global synchronization versus chimera behavior is controlled by the randomness of the initial state, as illustrated in Fig.~\ref{fig:snapshots}.

\subsection{Role of SU(2) symmetry in stabilizing synchronization} \label{sec:aniso}

\begin{figure}[htbp]
\includegraphics[width=0.48\textwidth]
{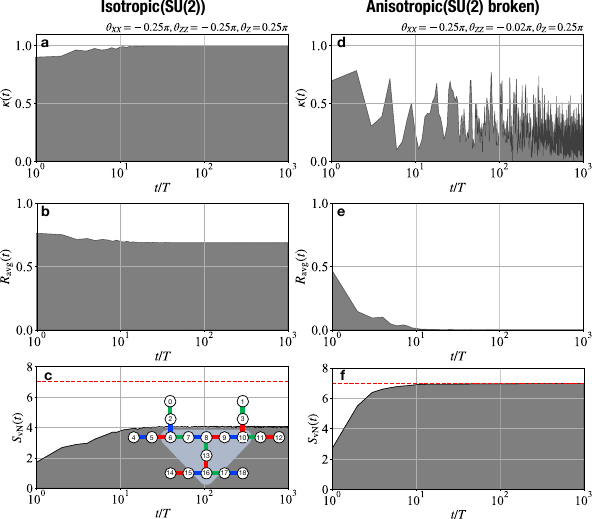}
\caption{
    \textbf{Effect of SU(2) symmetry breaking on synchronization and entanglement.} 
    Statevector simulations of long-time dynamics on a $L=19$ heavy-hex lattice. 
    Panels show the time evolution of the synchronization order parameter $\kappa(t)$, the mean oscillation radius $R_{\rm avg}(t)$, and the von Neumann entanglement entropy $S_{\rm vN}(t)$. 
    \textbf{a--c,} SU(2)-symmetric (isotropic) case with $\theta_{XX}=\theta_{ZZ}=-0.25\pi$. 
    \textbf{d--f,} SU(2)-broken (anisotropic) case with $(\theta_{XX},\theta_{ZZ})=(-0.25\pi,-0.02\pi)$. 
    The magnetic-field parameter is fixed to $\theta_Z=0.25\pi$, and the initial state is prepared with weak randomness $\phi^{\rm max}=\pi$. 
    The heavy-hex geometry used in the simulations is shown as an inset in panel \textbf{c}. 
    The entanglement entropy $S_{\rm vN}(t)$ is evaluated for the subsystem $A=\{6,7,8,9,10,13,16\}$ highlighted in the inset of panel \textbf{c}; the red dashed line indicates the maximal value $S_{\rm vN}=|A|=7$.
}
\label{fig:heavyhex19aniso}
\end{figure}

To examine the role of SU(2) symmetry in stabilizing synchronization, we introduce interaction anisotropy, $J_x\neq J_z$ (equivalently $\theta_{XX}\neq \theta_{ZZ}$), which explicitly breaks SU(2) invariance. 
We analyze long-time dynamics on a heavy-hex lattice with $L=19$ qubits using statevector simulations.
Figure~\ref{fig:heavyhex19aniso} compares isotropic and anisotropic cases for weak randomness ($\phi^{\max}=\pi$), focusing on the synchronization order parameter $\kappa(t)$, the average oscillation amplitude $R_{\rm avg}(t)$, and the von Neumann entanglement entropy $S_\text{vN}(t) = -\mathrm{Tr} \left[\rho_A(t) \log_2 \rho_A(t)\right]$ of a fixed subsystem $A$ (see Supplementary Information Sec.~S4.4 for details).

For isotropic interactions ($\theta_{XX}=\theta_{ZZ}$), $\kappa(t)$ rapidly approaches unity and $R_{\rm avg}(t)$ remains finite over long times, indicating persistent synchronized oscillations, while $S_{\rm vN}(t)$ grows and saturates at approximately half of its maximal value. 
In contrast, for anisotropic interactions, $\kappa(t)$ fails to converge to unity and $R_{\rm avg}(t)$ decays rapidly toward zero, while $S_{\rm vN}(t)$ approaches its maximal value, consistent with the loss of coherent oscillations and evolution toward a highly entangled state.
These trends are robust to variations in $\phi^{\max}$ and $(\theta_{XX},\theta_{ZZ})$, suggesting that the destabilization of synchronization is a generic consequence of SU(2) symmetry breaking (Supplementary Information Sec.~S4.4).

These results show that SU(2) symmetry stabilizes the observed collective oscillations: synchronization and chimera dynamics occur in the isotropic model but are destabilized once anisotropy breaks the symmetry (see Supplementary Information Sec.~S7).

\section{Discussion}

We have demonstrated two forms of collective nonequilibrium order on a programmable quantum processor: symmetry-protected synchronization and a quantum chimera state. 
By implementing Floquet dynamics of the isotropic Heisenberg model on IBM heavy-hex processors, we observed the self-organization of initially phase-randomized spins into coherent transverse-magnetization oscillations on $L=28$ qubits, and identified a device-scale regime on $L=156$ qubits where locally synchronized and unsynchronized regions coexist under homogeneous dynamics. 
Statevector simulations show that synchronization becomes nearly perfect in the absence of noise, while MPS calculations reproduce the spatial coexistence observed at large system size, indicating that the chimera behavior arises from the underlying coherent dynamics of the closed quantum system rather than experimental imperfections.

Our results identify SU(2) symmetry as the organizing principle stabilizing the synchronized response. 
Within the isotropic manifold, phase coherence and oscillation amplitude remain robust over long times, whereas explicit symmetry breaking through interaction anisotropy rapidly suppresses coherent oscillations and drives the system toward highly entangled states. 
This behavior can be understood from the structure of SU(2) multiplets, which enforce rigid level spacings and bundle many microscopic transitions into a common harmonic response. 
In this sense, the collective oscillations observed here constitute a symmetry-protected form of dynamical order: they persist under SU(2)-invariant dynamics but destabilize once the symmetry constraint is lifted.

The large-system regime reveals a quantum analogue of classical chimera states. 
In this regime, coherent and incoherent regions coexist in space despite homogeneous interactions and driving. 
The transition from global synchronization to spatial coexistence is controlled by the randomness of the initial state, indicating that the degree of initial phase disorder acts as a key parameter shaping the emergent dynamics. 
The persistence of locally synchronized pockets even in strongly randomized conditions further suggests that partial coherence can survive deep in regimes where global order is absent. 
Moreover, regions that appear incoherent at large scales can exhibit hierarchical structure: upon finer spatial inspection, smaller synchronized domains can emerge inside them, consistent with the nested organization characteristic of chimera dynamics.

More broadly, these results highlight how near-term quantum processors can probe nonequilibrium quantum many-body dynamics in two dimensions. 
While classical simulations remain essential for validating microscopic mechanisms and providing noise-free references, experiments access device-scale coherent dynamics that are increasingly difficult for classical methods to reach at comparable sizes and times. 
The combined strategy demonstrated here--quantum hardware together with noiseless numerical simulations--opens a path toward systematic exploration of emergent dynamical phenomena in interacting quantum systems.

\section{Methods}

\subsection{Model}

We study an $L$-qubit spin-$1/2$ system governed by the Hamiltonian
\begin{align}
\hat H &= \hat H_{\rm int} + \hat H_Z,
\end{align}
with
\begin{align}
\hat H_{\rm int} &= \sum_{\langle i,j\rangle}\hat H_{XXZ}^{(ij)},\qquad
\hat H_Z = h_z\sum_{i=0}^{L-1}\hat Z_i,
\end{align}
where
\begin{equation}
\hat H_{XXZ}^{(ij)}=-J_x(\hat X_i\hat X_j+\hat Y_i\hat Y_j)-J_z\hat Z_i\hat Z_j.
\end{equation}
Here $\hat X_i$, $\hat Y_i$, and $\hat Z_i$ denote Pauli operators acting on qubit $i$, and $|0_i\rangle$ and $|1_i\rangle$ are eigenstates of $\hat Z_i$ satisfying $\hat Z_i|0_i\rangle=|0_i\rangle$ and $\hat Z_i|1_i\rangle=-|1_i\rangle$. 
In the isotropic limit $J_x=J_z\equiv J$, the interaction reduces to the isotropic Heisenberg coupling 
\begin{equation}
\hat H_{XXX}^{(ij)}=-J(\hat X_i\hat X_j+\hat Y_i\hat Y_j+\hat Z_i\hat Z_j).
\end{equation}
Throughout this work, we focus on the ferromagnetic case $J>0$.
For $J<0$, the SU(2) symmetry remains intact and therefore the system is expected to exhibit similar synchronization behavior.

\begin{figure}[htbp]
\includegraphics[width=0.48\textwidth]{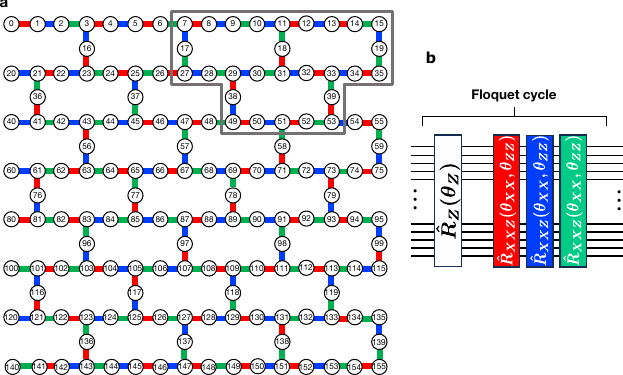}
\caption{
	\textbf{a,} Two-qubit connectivity of the \texttt{ibm\_kobe} processor forming a heavy-hex lattice with $L=156$ qubits.
	Circles denote qubits and edges denote native couplings. 
 	Three parallel layers of $\hat R_{XXZ}$ gates applied within a single Floquet cycle are highlighted in red, blue, and green.
	The gray boundary indicates the $L=28$-qubit subsystem used in the $L=28$ experiments. 
	\textbf{b,} Schematic circuit of a single Floquet cycle implementing the unitary operator $\hat U_{\rm F}$. 
	Colored boxes denote the three $\hat R_{XXZ}$ layers and white boxes denote single-qubit $\hat R_Z$ rotations.
}
\label{fig:circuit}
\end{figure}

\subsection{Floquet circuit}

We implement stroboscopic dynamics at discrete times $t=nT$ according to
\begin{equation}
\hat U(t=nT)=
\left[\left(\prod_{\langle i,j\rangle}e^{-i\hat H_{XXZ}^{(ij)}T}\right)e^{-i\hat H_Z T}\right]^n
\equiv \hat U_{\rm F}^n,
\end{equation}
where $\hat U_{\rm F}$ denotes the Floquet unitary for a single cycle of duration $T$. 
A single Floquet cycle is parametrized as
\begin{equation}
\hat U_{\rm F}=
\left[\prod_{\langle i,j\rangle}\hat R_{XXZ(i,j)}(\theta_{XX},\theta_{ZZ})\right]
\left[\prod_i \hat R_{Z_i}(\theta_Z)\right],
\end{equation}
with $\hat R_{Z_i}(\theta_Z)=\exp[-i(\theta_Z/2)\hat Z_i]$, $\theta_Z=h_zT$, and
\begin{equation}
\hat R_{XXZ(i,j)}(\theta_{XX},\theta_{ZZ})=
\exp\!\left[-\frac{i}{2}\left(\theta_{XX}\hat X_i\hat X_j+\theta_{XX}\hat Y_i\hat Y_j+\theta_{ZZ}\hat Z_i\hat Z_j\right)\right],
\end{equation}
where $\theta_{XX}=\theta_{YY}=-J_xT$ and $\theta_{ZZ}=-J_zT$.
For the isotropic implementation ($\theta_{XX}=\theta_{ZZ}$, corresponding to $J_x=J_z$), the SU(2) symmetry of the Floquet dynamics (in the absence of the longitudinal field) is discussed in Supplementary Information Sec.~S7. 
On the heavy-hex connectivity, the two-qubit gates are applied in three parallel layers per Floquet cycle (see Fig.~\ref{fig:circuit}).
Details of circuit compilation and device implementation are provided in Supplementary Information Sec.~S2.

\subsection{Quantum processor and measurements}

Experiments were performed on the IBM Heron-family processor \texttt{ibm\_kobe} accessed through the IBM Quantum cloud (June--July 2025). 
The native two-qubit entangling gate of the device is the CZ gate. 
For each stroboscopic time step, expectation values were estimated from $N_{\rm shots}=2^{13}$ projective measurements per circuit. 
Local observables $\langle \hat X_j(t)\rangle$ were obtained by measuring in the $X$ basis, implemented via single-qubit basis rotations followed by readout in the computational basis. 
A summary of the device calibration and operating conditions is provided in Supplementary Information Sec.~S2.
No additional error mitigation was applied beyond the normalization procedure described below.

\subsection{Initial states}

Initial states are product states with local Bloch vectors randomized in the $xy$ plane.
For each qubit $j$, the local state is prepared as 
\begin{equation}
|\psi_j(0)\rangle = \hat R_{Z_j}(\phi_j)\,\hat H_j\,|0_j\rangle,
\end{equation}
where $\hat H_j$ denotes the Hadamard gate acting on qubit $j$. 
The initial state of the entire system at $t=0$ is then given by $|\psi(0)\rangle=\bigotimes_j|\psi_j(0)\rangle$.
The phases $\phi_j$ are sampled independently and uniformly from $[0,\phi^{\max}]$, where $\phi^{\max}$ controls the degree of phase randomness in the initial state. 
Throughout this work, we mainly consider $\phi^{\max}=\pi$ and $2\pi$.

\subsection{Error mitigation}

To compensate for circuit-depth-dependent attenuation of local traceless observables on hardware, we apply a normalization procedure motivated by a global depolarizing noise model~\cite{Swingle2018,Vovrosh2021,Urbanek2021} (Supplementary Information Sec.~S3). 
Raw (unmitigated) device estimates are denoted by $\langle\cdot\rangle_0$. 
At each stroboscopic time $t_n=nT$, we estimate a normalization factor using a reference circuit in which the ideal average magnetization has unit magnitude (ferromagnetic initial state $\{\phi_j=0\}$ and $\theta_Z=0$): 
\begin{equation}
f(t_n)=\Bigl|\bigl\langle \avg{X}(t_n)\bigr\rangle_{0,\{\phi_j=0\},\theta_Z=0}\Bigr|,
\qquad
\avg{X}(t)=\frac{1}{|M|}\sum_{j\in M}\hat X_j(t).
\end{equation}
Here $M$ denotes the set of measured qubits (typically all qubits in the implemented $L=28$ or $L=156$ system).
The target data at the same circuit depth are then normalized as 
\begin{equation}
\langle \hat X_j(t_n)\rangle \approx \frac{\langle \hat X_j(t_n)\rangle_0}{f(t_n)}.
\label{eq:method-mitigation}
\end{equation}
This procedure primarily compensates the dominant multiplicative attenuation of traceless observables; residual coherent errors may remain (see Supplementary Information Sec.~S3).
Uncertainty bars are obtained by propagating the finite-shot uncertainties of the numerator and denominator through Eq.~(\ref{eq:method-mitigation}). 
Validation of the normalization-based mitigation for $L=28$ and $L=156$ is shown in Supplementary Figs.~S2 and S3.
No additional error-suppression techniques (such as dynamical decoupling~\cite{Viola1999,Souza2012}) or error-mitigation methods (such as zero-noise extrapolation~\cite{Temme2017,Li2017} or probabilistic error cancellation~\cite{van_den_Berg2023}) were used in the results reported here.

\subsection{Synchronization order parameter}

In statevector and MPS simulations, where both $\langle \hat X_j(t)\rangle$ and $\langle \hat Y_j(t)\rangle$ are available, we compute the Kuramoto-type order parameter
\begin{equation}
\kappa(t)=\left|\frac{1}{|M|}\sum_{j\in M}e^{i\Theta_j(t)}\right|,
\end{equation}
where the phases $\Theta_j(t)$ are extracted from the complex local magnetization 
$w_j(t)=\langle \hat X_j(t)\rangle+i\langle \hat Y_j(t)\rangle = R_j(t)e^{i\Theta_j(t)}$.

On hardware, measuring both $\langle \hat X_j(t)\rangle$ and $\langle \hat Y_j(t)\rangle$ would double the measurement overhead. 
Instead, we employ a proxy order parameter $\tilde{\kappa}(t)$ obtained from phases extracted from the analytic signal of the real-valued time series $x_j(t_n)=\langle \hat X_j(t_n)\rangle$. 
The analytic signal is constructed via a discrete Hilbert-transform construction~\cite{Gabor1946, Pikovsky2001}, from which the instantaneous phase $\tilde{\Theta}_j(t)$ is obtained. 
As a proxy for $w_j(t)$, we define $\tilde w_j(t)=x_j(t)+i{\cal H}[x_j](t)=\tilde R_j(t)e^{i\tilde\Theta_j(t)}$, where ${\cal H}[x_j](t)$ denotes the Hilbert transform of $x_j(t)$. 
The proxy order parameter is then defined as
\begin{equation}
\tilde{\kappa}(t)=\left|\frac{1}{|M|}\sum_{j\in M}e^{i\tilde{\Theta}_j(t)}\right|.
\end{equation}
Definitions and benchmarking of this proxy are provided in Supplementary Information Sec.~S1. 
We validate this proxy by comparing it with the standard order parameter $\kappa(t)$ computed from $w_j(t)$ in MPS simulations;
the two quantities show closely matched behavior over the full time window studied (Supplementary Fig.~S1).

For spatial maps on large lattices, we define a local order parameter centered on qubit $j$,
\begin{equation}
\tilde{\kappa}_{j}(t)
=
\left|
\frac{1}{|\mathcal{N}_{j}^{(K)}|}
\sum_{\ell\in \mathcal{N}_{j}^{(K)}}
e^{i\tilde{\Theta}_{\ell}(t)}
\right|,
\label{eq:kappa_local}
\end{equation}
where $\mathcal{N}_{j}^{(K)}$ denotes the set of qubits whose shortest-path distance from $j$ on the heavy-hex graph is at most $K$, including $j$ itself.
The phases $\tilde{\Theta}_{\ell}(t)$ are obtained from $\langle \hat X_{\ell}(t)\rangle$ via the discrete Hilbert-transform procedure described above.
In the main text, we use $K=4$, which corresponds to neighborhoods containing about $16$ qubits on average, and show late-time temporal averages $\langle \tilde{\kappa}_{j}(t)\rangle_{\mathcal{T}}$ over the time interval $\mathcal{T}=[25,30]$.
The corresponding quantity $\kappa_{j}(t)$ is defined by replacing $\tilde{\Theta}_{\ell}(t)$ with $\Theta_{\ell}(t)$ in Eq.~(\ref{eq:kappa_local}).
Spatial maps obtained for other graph-distance cutoffs are presented in Supplementary Information Sec.~S6.3.

\subsection{Classical simulations}

For smaller systems, including $L=28$, we perform exact statevector simulations of the Floquet dynamics under $\hat U_{\rm F}$ to obtain $\langle \hat X_j(t)\rangle$ and $\langle \hat Y_j(t)\rangle$. 
For larger systems (notably $L=156$), we simulate the dynamics using MPS time evolution with bond dimension $\chi$. 
Unless otherwise stated, results are shown for $\chi=600$, and convergence checks are reported in Supplementary Information Sec.~S5.

\subsection{Parameter choices}

Unless otherwise stated, the parameters are set to $(\theta_{XX},\theta_{ZZ},\theta_Z)=(-0.25\pi,-0.25\pi,0.25\pi)$ for the SU(2)-symmetric case. 
For the anisotropy study, $\theta_{XX}$ and $\theta_{ZZ}$ are varied while keeping $\theta_Z=0.25\pi$ fixed. 
The parameter $\theta_Z$ sets the dominant precession scale of the transverse magnetization; therefore changing $\theta_Z$ modifies the observed oscillation period in stroboscopic time.


\begin{acknowledgments}
This work was supported in part by the New Energy and Industrial Technology Development Organization (NEDO), Japan (Project No. JPNP20017). 
We acknowledge support by the Japan Society for the Promotion of Science (JSPS), KAKENHI (Grant Nos.
JP21H04446, 
JP22K03520, and
JP23K13066) 
from the Ministry of Education, Culture, Sports, Science and Technology (MEXT), Japan. 
We also thank the Japan Science and Technology Agency (JST) for support through COI-NEXT (Grant No. JPMJPF2221) and MEXT for the Program for Promoting Research of the Supercomputer Fugaku (Grant No. MXP1020230411).  
Further support was provided by the UTokyo Quantum Initiative, 
the RIKEN TRIP initiative (RIKEN Quantum and Many-Body Electron Systems), and the Center of Excellence (COE) Research Grant in Computational Science from Hyogo Prefecture and Kobe City through the Foundation for Computational Science.
Part of the numerical simulations was carried out on the HOKUSAI supercomputer at RIKEN and the supercomputer system at the D3 center, Osaka University, through the HPCI System Research Project (Project ID: hp250062).
The MPS simulations were performed using the ITensor library~\cite{Fishman2022}. 
\end{acknowledgments}

\bibliography{bibdtc}

\end{document}